\documentclass[conference]{IEEEtran}
\IEEEoverridecommandlockouts
\usepackage{cite}
\usepackage{amsmath,amssymb,amsfonts}
\usepackage{algorithmic}
\usepackage{graphicx}
\usepackage{textcomp}
\usepackage{xcolor}
\usepackage{bm}
\usepackage{threeparttable}
\usepackage{multirow}
\usepackage{booktabs}
\usepackage{makecell}
\usepackage{orcidlink}
\usepackage{threeparttable}

\usepackage{lipsum}
\usepackage{fancyhdr}
\fancypagestyle{sec}{\lhead{\tmpx}}

\fancypagestyle{arXiv}
{
   \fancyhf{}
   \chead{\textcolor{gray}{This article will be presented at the Asia and South Pacific Design Automation Conference (ASP-DAC 2026), Hong Kong, China, January 19–22, 2026.}}
   \cfoot{ \fontsize{=9pt}{10pt}\selectfont  \textcolor{gray}{© 2026 IEEE. Personal use of this material is permitted. Permission from IEEE must be obtained for all other uses, in any current or future media, including reprinting/republishing this material for advertising or promotional purposes, creating new collective works, for resale or redistribution to servers or lists, or reuse of any copyrighted component of this work in other works}\\\fontsize{=10pt}{12pt}\selectfont\thepage}
   
}

\def\BibTeX{{\rm B\kern-.05em{\sc i\kern-.025em b}\kern-.08em
    T\kern-.1667em\lower.7ex\hbox{E}\kern-.125emX}}
\begin{document}

\title{JaneEye: A 12-nm 2K-FPS 18.9-$\mu$J/Frame Event-based Eye Tracking Accelerator\\}
\author{
Tao Han\textsuperscript{1,*}\orcidlink{0009-0002-5661-985X}, 
Ang Li\textsuperscript{1,*}\orcidlink{0000-0003-3615-6755}, 
Qinyu Chen\textsuperscript{2}\orcidlink{0009-0005-9480-6164}, 
Chang Gao\textsuperscript{1}\orcidlink{0000-0002-3284-4078}\\
\IEEEauthorblockA{\textsuperscript{1}Department of Microelectronics, Delft University of Technology, The Netherlands \\
\textsuperscript{2}Leiden Institute of Advanced Computer Science (LIACS), Leiden University, The Netherlands}
\thanks{{\textsuperscript{*}}Equal Contribution.}
\thanks{
Corresponding authors: Chang Gao (chang.gao@tudelft.nl) and Qinyu Chen (q.chen@liacs.leidenuniv.nl).
}
}

\maketitle

\begin{abstract}
Eye tracking has become a key technology for gaze-based interactions in Extended Reality (XR). However, conventional frame-based eye-tracking systems often fall short of XR's stringent requirements for high accuracy, low latency, and energy efficiency. Event cameras present a compelling alternative, offering ultra-high temporal resolution and low power consumption.
In this paper, we present JaneEye, an energy-efficient event-based eye-tracking hardware accelerator designed specifically for wearable devices, leveraging sparse, high-temporal-resolution event data. We introduce an ultra-lightweight neural network architecture featuring a novel ConvJANET layer, which simplifies the traditional ConvLSTM by retaining only the forget gate, thereby halving computational complexity without sacrificing temporal modeling capability. Our proposed model achieves high accuracy with a pixel error of 2.45 on the 3ET+ dataset, using only 17.6K parameters, with up to 1250\,Hz event frame rate. To further enhance hardware efficiency, we employ custom linear approximations of activation functions (HardSigmoid and HardTanh) and fixed-point quantization. Through software-hardware co-design, our 12-nm ASIC implementation operates at 400 MHz, delivering an end-to-end latency of 0.5 ms (equivalent to 2000 Frames Per Second (FPS)) at an energy efficiency of 18.9 $\mu$J/frame. JaneEye sets a new benchmark in low-power, high-performance eye-tracking solutions suitable for integration into next-generation XR wearables.
\end{abstract}


\section{Introduction}
\thispagestyle{arXiv}
Extended Reality (XR) is rapidly reshaping how people perceive and engage with digital environments. Eye tracking, which monitors and records eye movements, has become crucial for creating immersive XR experiences~\cite{plopski2022eye}, particularly following the launch of the Apple Vision Pro in June 2023. By enabling gaze-based interactions~\cite{fernandes2023leveling,hu2022gaze,ding2024facet}, eye tracking allows users to navigate and control virtual spaces simply by looking. While significant progress has been made toward integrating this technology into wearable devices~\cite{menendez2024eye}, challenges such as latency and power consumption still need to be addressed to deliver a smooth user experience~\cite{fernandes2023leveling}.

The human eye is the fastest-moving organ, capable of movements exceeding 300°/s~\cite{ramachandran2002encyclopedia}. Eye tracking requires extremely high sampling rates (in the kilohertz range) to capture rapid eye movements, ensuring smooth tracking and reducing motion sickness in virtual environments. To be integrated into mobile devices, eye tracking systems must be extremely power-efficient and lightweight, accommodating limited battery capacity and compact form factors. However, current head-mounted devices (HMDs) with traditional frame-based eye-tracking systems consume a significant amount of energy when achieving kilohertz frame rates. They capture full images at fixed intervals regardless of scene changes, meaning each frame involves reading data from all pixels. The large data volumes require high bandwidth and significant energy for transfer and processing, posing challenges for real-time applications on wearable devices. A recent study reports tracking delays between 45 and 81\,ms in various HMD eye trackers~\cite{stein2021comparison}, falling short of kilohertz frame rates.

Event cameras, also known as Dynamic Vision Sensors (DVS)~\cite{li2015design,gallego2020event,lenero20113}, offer a powerful alternative for addressing eye tracking challenges. By capturing only brightness changes, event cameras generate sparse, asynchronous events that offer high temporal resolution and low power consumption. This sensing mechanism produces less data and reduces processing demands during fixation while accurately capturing fast eye movements during saccades. Event cameras offer a significant advantage: their sparse output and dynamic event rate can substantially reduce energy consumption in eye-tracking systems. Additionally, their detection principle enables much higher maximum sampling rates than traditional cameras.

Deep learning has become a promising approach for event-based eye tracking. Traditional CNN-based architectures were proposed in~\cite{li2023track, li2024gaze}. However, extracting only spatial features proves insufficient for accurate eye tracking. Therefore, spatiotemporal models were introduced to extract information both spatially and temporally~\cite{chen20233et,pei2024lightweight,wang2024mambapupil,wu2025brat}. For example, 3ET~\cite{chen20233et} adopts a CONV-LSTM, MambaPupil~\cite{wang2024mambapupil} leverages a bidirectional Gated Recurrent Unit (GRU) and a linear time-varying State Space Module (SSM), and BRAT~\cite{wu2025brat} uses a bidirectional relative positional attention transformer. While these neural networks have achieved impressive prediction accuracy, their efficiency remains a significant challenge requiring further investigation.

Beyond eye-tracking algorithms, designing dedicated hardware capable of achieving high frame rates (over 1 kHz) while maintaining low power consumption (at the milliwatt level) poses a major challenge. Currently, research on hardware design for eye tracking systems is limited. The ASIC designs proposed in~\cite{tan2025toward} and~\cite{zhao2022flatcam} focused on gaze estimation, differing from our pupil detection task. Zhang et al. proposed an FPGA-based pupil detection system utilizing Submanifold Sparse CNN (SCNN) in~\cite{zhang2024co}. However, the system demonstrated limited accuracy, with an average error of 3.71 pixels, and suffered from high power consumption at the watt level. Retina~\cite{bonazzi2024retina} achieved minimal power consumption, but at the expense of compromised frame rates and accuracy.

In this paper, we propose JaneEye, an energy-efficient event-based eye-tracking hardware accelerator specifically designed for wearable XR devices. We design an ultra-compact neural network with the novel ConvJANET layer, which simplifies ConvLSTM by retaining only the forget gate, reducing computational complexity by 50\% while maintaining temporal modeling accuracy. Our approach achieves high prediction accuracy (2.45 pixel error on the 3ET+ dataset) using only 17.6K parameters, with up to 1250\,Hz frame rate. We implement custom linear approximations for activation functions and fixed-point quantization to maximize efficiency. Our 12-nm ASIC implementation achieves 0.5 ms end-to-end latency (2000 FPS) and 18.9~$\mu$J/frame energy consumption, setting a new benchmark for low-power, high-speed eye tracking.
\begin{figure*}[t]
    \centering    \includegraphics[width=\textwidth]{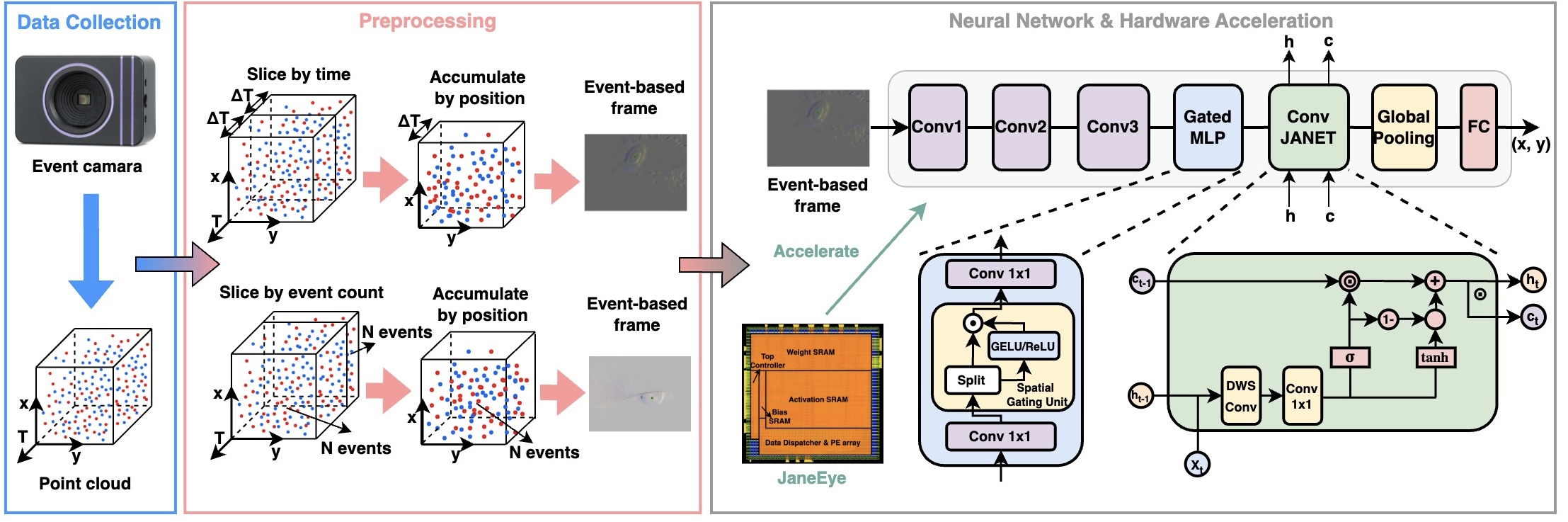}
    \caption{End-to-end flowchart of the proposed JaneEye eye-tracking system. The pipeline consists of three main stages. (1) Data Collection: An event camera captures sparse spatiotemporal data, generating a point cloud of events. (2) Preprocessing: This asynchronous point cloud is converted into dense 2D 'event-based frames' using two alternative aggregation methods: 'Slice by time' ($\Delta T$) or 'Slice by event count' ($N$ events). (3) JaneEye-Net Neural Network \& JaneEye Hardware Acceleration: The resulting frame is fed into the lightweight JaneEye-Net, which uses three convolutional layers (Conv1-3) for spatial feature extraction, a Gated MLP, and our novel ConvJANET layer for spatiotemporal modeling. Finally, a Global Pooling and Fully Connected (FC) layer regress the (x, y) pupil coordinates. The JaneEye ASIC accelerates the JaneEye-Net eye-tracking neural network.}
    \label{neural_network}
\end{figure*}
\begin{figure}[t] 
\centering
\includegraphics[width=1.0\linewidth]{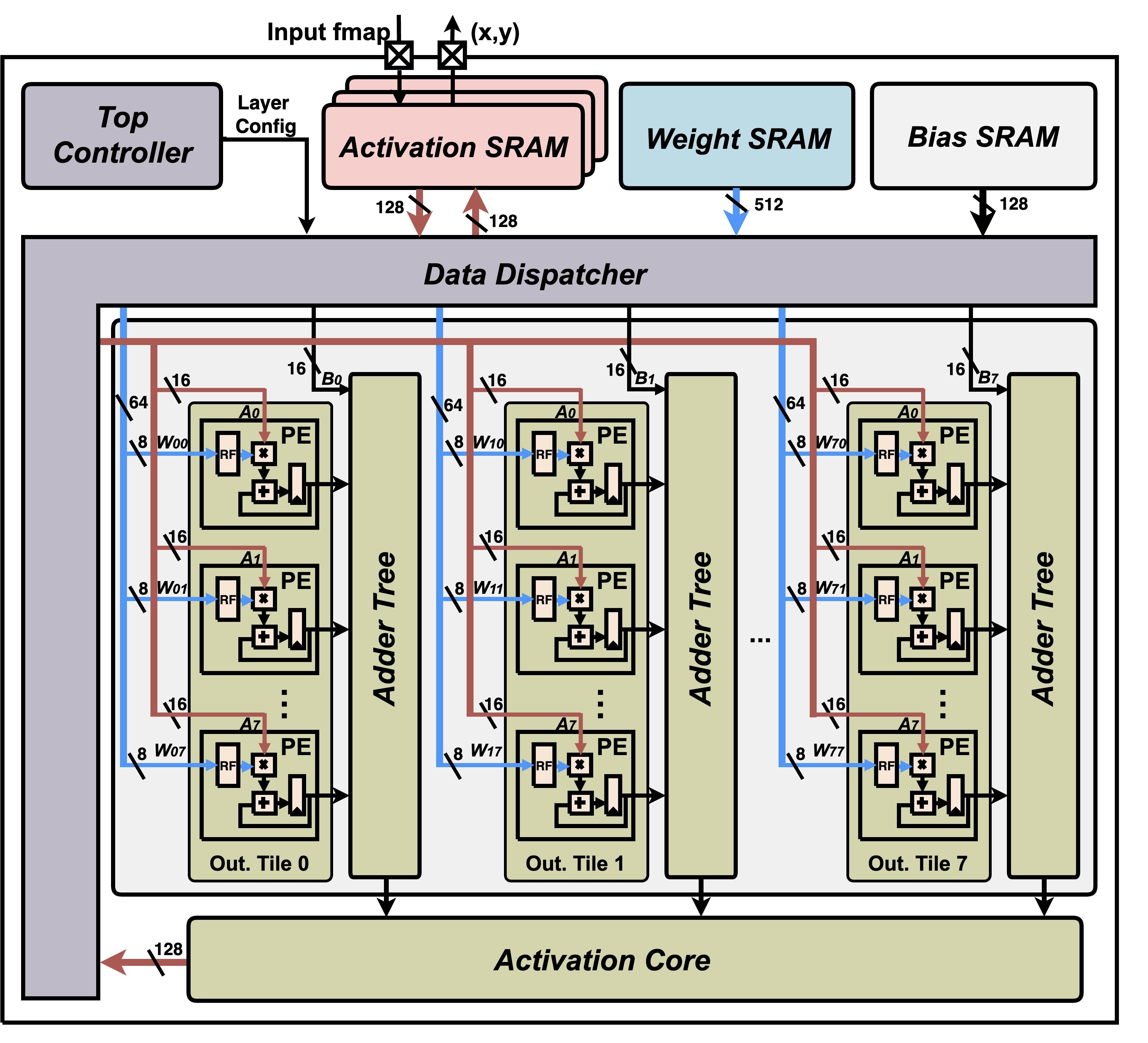}
\caption{Microarchitecture of the proposed JaneEye hardware accelerator. The design is managed by a Top Controller and features dedicated on-chip SRAMs for Activations, Weights, and Biases to support high-bandwidth parallel memory access. A Data Dispatcher broadcasts data to the main computational core, which is organized as an array of 8 parallel Output Tiles. Each tile contains 8 PEs, and their partial sum outputs are aggregated by a dedicated Adder Tree. This 64-PE (8$\times$8) array performs the bulk of the MAC operations. The final results are passed to an Activation Core for nonlinear function processing.}
\label{fig:hardware_architecture}
\end{figure}
\begin{figure}[t] 
\centering
\includegraphics[width=1.0\linewidth]{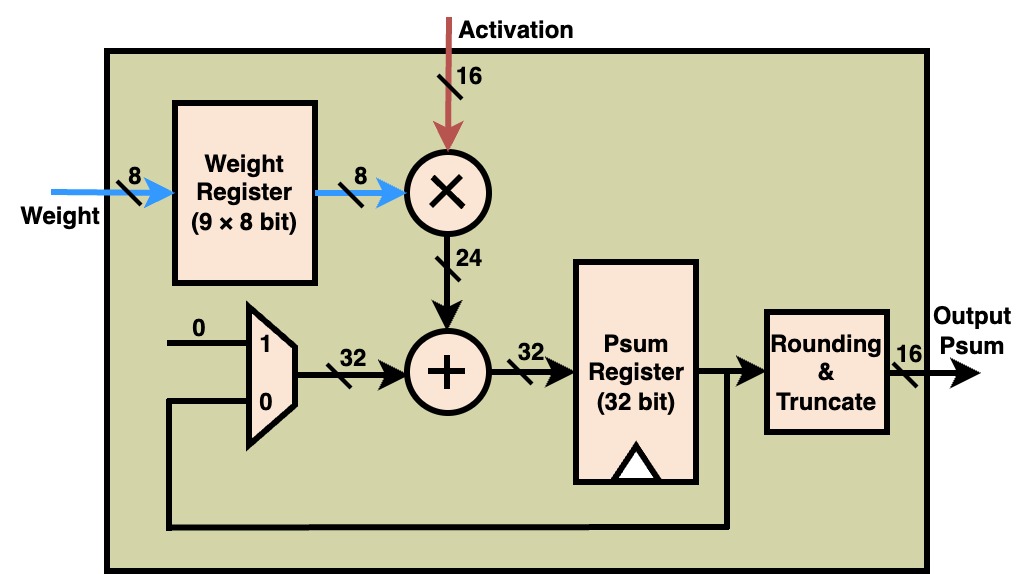}
\caption{Detailed architecture of a single PE for MAC operations. It features a 9 $\times$ 8-bit Weight Register, which allows for storing and reusing an entire 3$\times$3 convolution kernel locally, minimizing data movement. In a processing cycle, an 8-bit weight from the register is multiplied with a 16-bit activation. The 24-bit result is fed to a 32-bit adder. A multiplexer (MUX) selects whether to add this product to the previously accumulated value from the 32-bit Psum Register (accumulation step) or to '0' (to start a new computation). The final 32-bit partial sum (Psum) is passed to a Rounding \& Truncate unit to produce the 16-bit output.}
\label{fig:PE_architecture}
\end{figure}
\section{Algorithm Design}
\subsection{Event-to-Frame Representation}
Event cameras capture asynchronous brightness changes as a stream of events $\mathcal{E} = \{e_i\}_{i=1}^{N}$, where each event $e_i = (t_i, x_i, y_i, p_i)$ consists of a timestamp $t_i$, pixel coordinates $(x_i, y_i)$, and polarity $p_i \in \{-1, +1\}$ indicating brightness decrease or increase, respectively. To leverage existing deep learning architectures, we convert this asynchronous event stream into frame-based representations.

We employ two complementary event aggregation strategies:

\subsubsection{Time-Based Event Aggregation}
For a given time window $[t_k, t_k + \Delta T]$, we construct a frame $\mathbf{F}_k^{time} \in \mathbb{R}^{H \times W}$ by accumulating all events within this interval:

\begin{equation}
\mathbf{F}_k^{time}(x, y) = \sum_{e_i \in \mathcal{E}_k} p_i \cdot \delta(x - x_i, y - y_i)
\label{eq:time_based_frame}
\end{equation}

where $\mathcal{E}_k = \{e_i : t_k < t_i \leq t_k + \Delta T\}$ represents the set of events in the $k$-th time window, and $\delta(\cdot, \cdot)$ is the Kronecker delta function. We set $\Delta T = 10$ ms to strike a balance between temporal resolution and computational efficiency.

\subsubsection{Event Count-Based Aggregation}
Alternatively, we aggregate a fixed number of events $N_{evt}$ to form each frame:

\begin{equation}
\mathbf{F}_k^{count}(x, y) = \sum_{i=(k-1)N_{evt}+1}^{kN_{evt}} p_i \cdot \delta(x - x_i, y - y_i)
\label{eq:count_based_frame}
\end{equation}

This approach adapts the temporal resolution to the scene dynamics, with higher frame rates during rapid eye movements. We empirically set $N_{evt} = 5000$ events per frame.

Following frame construction, we apply spatial downsampling by a factor of 8 using bilinear interpolation, reducing the spatial resolution from $640 \times 480$ to $80 \times 60$ pixels. This preprocessing step reduces computational requirements while preserving essential spatial information for pupil localization.

\subsection{JaneEye Network Architecture}

Inspired by recent advances in efficient neural network architectures evaluated on the eye tracking task~\cite{wang2024event,chen2025event}, we propose JaneEye-Net, an ultra-lightweight neural network specifically designed for event-based eye tracking. The architecture, illustrated in Fig.~\ref{neural_network}, consists of four main components:

\subsubsection{Spatial Feature Extraction}
The network begins with a three-layer convolutional backbone that progressively extracts hierarchical spatial features:

\begin{equation}
\begin{aligned}
\mathbf{h}_1 &= \text{ReLU}(\text{Conv}_{7\times7}(\mathbf{F}_k)) \\
\mathbf{h}_2 &= \text{ReLU}(\text{Conv}_{3\times3}(\mathbf{h}_1)) \\
\mathbf{h}_3 &= \text{ReLU}(\text{Conv}_{3\times3}(\mathbf{h}_2))
\end{aligned}
\label{eq:conv_layers}
\end{equation}

where $\text{Conv}_{k\times k}$ denotes a 2D convolution with kernel size $k \times k$. The decreasing kernel sizes (7$\times$7, 3$\times$3, 3$\times$3) enable the network to capture both large-scale eye structure and fine-grained pupil details.

\subsubsection{Gated Multilayer Perceptron (GMLP)}
Following~\cite{liu2021pay}, we incorporate a gated MLP layer to model spatial interactions efficiently. Given input features $\mathbf{X} \in \mathbb{R}^{C \times H \times W}$, the GMLP operates as:

\begin{equation}
\begin{aligned}
\mathbf{Z} &= \text{Conv}_{1\times1}(\mathbf{X}) \in \mathbb{R}^{2C \times H \times W} \\
\mathbf{Z}_1, \mathbf{Z}_2 &= \text{Split}(\mathbf{Z}) \\
\mathbf{Y} &= \mathbf{Z}_1 \odot \text{GELU}(\mathbf{Z}_2) \\
\text{GMLP}(\mathbf{X}) &= \text{Conv}_{1\times1}(\mathbf{Y})
\end{aligned}
\label{eq:gmlp}
\end{equation}

where $\odot$ denotes element-wise multiplication, and the split operation divides $\mathbf{Z}$ equally along the channel dimension.

\subsubsection{Convolutional JANET (ConvJANET)}
To capture temporal dependencies across event frames while maintaining computational efficiency, we introduce ConvJANET, a lightweight variant of ConvLSTM that retains only the forget gate~\cite{van2018unreasonable}. This design reduces parameters and computations by 50\% compared to standard ConvLSTM while preserving temporal modeling capability.

Given input $\mathbf{x}_t$ and previous hidden state $\mathbf{h}_{t-1}$, ConvJANET updates its states as:

\begin{equation}
\begin{aligned}
\mathbf{f}_t &= \sigma(\mathcal{F}_{conv}([\mathbf{x}_t, \mathbf{h}_{t-1}])) \\
\tilde{\mathbf{c}}_t &= \tanh(\mathcal{G}_{conv}([\mathbf{x}_t, \mathbf{h}_{t-1}])) \\
\mathbf{c}_t &= \mathbf{f}_t \odot \mathbf{c}_{t-1} + (1 - \mathbf{f}_t) \odot \tilde{\mathbf{c}}_t \\
\mathbf{h}_t &= \mathbf{c}_t
\end{aligned}
\label{eq:convjanet_formal}
\end{equation}

where $[\cdot, \cdot]$ denotes concatenation, $\mathcal{F}_{conv}$ and $\mathcal{G}_{conv}$ are learnable convolutional transformations implemented as depth-wise separable convolutions followed by $1\times1$ convolutions, $\sigma$ is the sigmoid function, and $\mathbf{f}_t$ represents the forget gate that controls information flow from previous time steps.

\subsubsection{Pupil Localization Head}
The final detection head employs global max pooling to extract the most salient features from each channel, followed by a fully connected layer that regresses the pupil center coordinates:

\begin{equation}
\begin{aligned}
\mathbf{g} &= \text{GlobalMaxPool}(\mathbf{h}_t) \\
(\hat{x}, \hat{y}) &= \text{FC}(\mathbf{g}) \in \mathbb{R}^{2}
\end{aligned}
\label{eq:detection_head}
\end{equation}

\subsection{Hardware-Aware Optimizations}

To enable efficient hardware implementation, we apply several co-design techniques that balance model accuracy with computational efficiency:

\subsubsection{Activation Function Approximation}
We replace computationally expensive nonlinear functions with piecewise linear approximations optimized for hardware implementation:

Customized HardSigmoid:
\begin{equation}
\sigma_{hard}(x) = \begin{cases}
0, & x < -4 \\
\frac{x}{8} + \frac{1}{2}, & -4 \leq x \leq 4 \\
1, & x > 4
\end{cases}
\label{eq:hardsigmoid_new}
\end{equation}

Customized HardTanh:
\begin{equation}
\tanh_{hard}(x) = \begin{cases}
-1, & x < -2 \\
\frac{x}{2}, & -2 \leq x \leq 2 \\
1, & x > 2
\end{cases}
\label{eq:hardtanh_new}
\end{equation}

These approximations enable implementation using only comparators and bit-shift operations, eliminating the need for complex arithmetic units. Additionally, we replace GELU with ReLU in the GMLP layer in the retraining phase discussed later for even more efficient hardware implementation.

\subsubsection{Mixed-Precision Quantization}
We employ a mixed-precision quantization scheme that balances memory footprint with numerical precision:
\begin{itemize}
    \item Weights: 8-bit fixed-point (Q1.7 format)
    \item Activations: 16-bit fixed-point (Q5.11 format)
\end{itemize}

This quantization strategy reduces memory requirements by 75\% for weights and 50\% for activations compared to 32-bit floating-point, while maintaining less than 1\% accuracy degradation.

\subsubsection{Retraining Strategy}
To mitigate the loss of accuracy due to hardware optimizations, we employ a progressive retraining approach. After each optimization step (activation approximation or quantization), we fine-tune the model for 10 epochs using the original training objective. This strategy successfully recovers performance, limiting total accuracy degradation to 1.6\% while achieving significant hardware efficiency gains.

\section{Hardware Accelerator Design}

Figure~\ref{fig:hardware_architecture} shows the microarchitecture of our hardware accelerator, designed to efficiently execute the JaneEye-Net model with minimal power consumption.

\subsection{System Architecture}

The accelerator consists of five main components. A top-level controller implemented as a 12-state FSM manages layer execution and dataflow reconfiguration. The memory subsystem includes three separate SRAMs: 64 KB for weights, 32 KB for activations, and 4 KB for biases. This separation enables three concurrent memory accesses per cycle, achieving an aggregate bandwidth of 3.2 GB/s at 400 MHz.

The data dispatcher implements double buffering with 16-entry FIFOs, hiding the 8-cycle SRAM read latency. The computational core contains 64 Processing Engines (PEs) arranged in an 8$\times$8 array, where each row shares input activations through a broadcast bus. The activation core implements four functions (bypass, ReLU, HardSigmoid, HardTanh) using comparators and bit shifters, requiring only 2 clock cycles per operation.

\subsection{Processing Engine Architecture}

Figure~\ref{fig:PE_architecture} details the PE microarchitecture. Each PE contains nine 8-bit weight registers and one 32-bit accumulator. The weight registers enable 9$\times$ reuse for 3$\times$3 convolutions without additional memory accesses. The 8-bit multiplier and 32-bit adder form a single-cycle MAC unit, with output truncation to 16 bits using convergent rounding.

The PE supports two dataflow modes via 2:1 multiplexers on all data paths. In weight-stationary mode, weights remain in registers for 49 cycles (7$\times$7 convolution) or 9 cycles (3$\times$3 convolution). In output-stationary mode, partial sums accumulate locally while weights and activations stream through. Mode switching requires 2 cycles for pipeline flushing.

\subsection{Dataflow Mapping}

Each layer type uses a specific dataflow pattern. Convolutional layers employ weight-stationary dataflow, achieving 98\% weight reuse for the 7$\times$7 layer and 89\% for the 3$\times$3 layers. The ConvJANET layer utilizes output-stationary dataflow, resulting in a 62\% reduction in partial sum writes compared to weight-stationary mapping. This reduction occurs because each PE accumulates 8 input channels before writing back. The fully connected layer uses row-stationary dataflow, processing 8 outputs in parallel across PE columns.
\begin{table}[tbp]
\centering
\setlength{\tabcolsep}{2pt}
\renewcommand{\arraystretch}{1.3}
\caption{Optimization ablation study in FP32 precision}
\label{table:optimization_ablation_study}
\resizebox{0.5\textwidth}{!}{%
\begin{threeparttable}
\begin{tabular}{cccccc}
\toprule
\textbf{Progressing Optimizations} & \makecell{\textbf{Params}\\\textbf{(K)}} & \makecell{\textbf{Model size}\\\textbf{reduction}} &
\makecell{\textbf{FLOPs}\\\textbf{(M)}} &
\makecell{\textbf{FLOPs}\\\textbf{reduction}} &
\makecell{\textbf{Precision}\\\textbf{(pixels)}} \\
\midrule
ERVT (baseline)\tnote{1} & 150 & - & 148 & - & 2.53\\
Single-stage architecture & 92 & 38.7\% & 92.0 & 37.8\% & 2.37\\
Adjust channel dimension & 45 & 70.0\% & 30.9 & 79.1\% & 2.62\\
Add global max pooling & 26 & 82.7\% & 30.8 & 79.2\% & 2.43  \\
Replace LSTM with ConvJANET & 22 & 85.3\% & 25.8 & 82.6\% & 2.41  \\
Replace attention with conv & 17.7 & 88.2\% & 10.7 & 92.8\% & 2.42  \\
Remove LayerNorm & 17.6 & 88.3\% & 10.7 & 92.8\% & 2.41  \\
\bottomrule
\end{tabular}
\begin{tablenotes}\footnotesize
\item[1] Our reimplementation of ERVT achieves 2.53 pixels on 3ET+, versus 2.48 pixels reported in the original paper, likely due to training details.
\end{tablenotes}
\end{threeparttable}
}
\end{table}
\begin{table}[tbp]
\centering
\setlength{\tabcolsep}{2pt}
\renewcommand{\arraystretch}{1.3}
\caption{Performance before/after applying software-hardware co-design techniques}
\label{table:software-hardware codesign}
\resizebox{0.5\textwidth}{!}{%
\begin{tabular}{cccc}
\toprule
\textbf{Step} & \makecell{\textbf{Precision}\\\textbf{(before retrain)}} & \makecell{\textbf{Precision}\\\textbf{(after retrain)}} & \makecell{\textbf{Precision drop}\\\textbf{(Pixel Error)}} \\
\midrule
Baseline  & \textbf{2.41} & N/A & N/A \\
GELU $\rightarrow$ ReLU & 6.28 & 2.42 & 0.01  \\
Sigmoid $\rightarrow$ Custom HardSigmoid & 8.10 & 2.43 & 0.02  \\
Tanh $\rightarrow$ Custom HardTanh & 7.12 & 2.43 & 0.02   \\
Quantization (W8A16) & \textbf{2.45} & N/A & \textbf{0.04}  \\
\bottomrule
\end{tabular}
}
\end{table}
\subsection{Memory Access Optimization}

Input feature maps are tiled into 8$\times$8 spatial blocks with 8 channels, matching the dimensions of the PE array. This tiling requires 4 KB of on-chip buffer storage and eliminates external memory access during tile computation. Each tile processes in 64 cycles for 3$\times$3 convolutions or 392 cycles for 7$\times$7 convolutions.

Zero-skipping logic detects zero activations using OR-trees and gates the corresponding MAC operations. Measurements on the 3ET+ dataset show 38-42\% activation sparsity, reducing dynamic power by 35\%. The prefetch unit begins loading the next tile at cycle 48 (for 3$\times$3) or cycle 376 (for 7$\times$7), achieving 94\% overlap between computation and memory access.

These optimizations enable an average PE utilization of 90\%, with individual layer utilization ranging from 87\% (ConvJANET) to 93\% (7$\times$7 convolution).

\section{Experimental Results}
\subsection{Experimental Setup}
\subsubsection{Software Configuration}
We evaluate our approach on the 3ET+ dataset~\cite{wang2024event,chen2025event}, a comprehensive 9.2 GB benchmark for event-based eye tracking. The dataset contains recordings from 13 subjects (2–6 sessions each) captured using a DVXplorer Mini event camera. It encompasses five distinct eye movement types: random drift, saccades, reading, smooth pursuits, and blinks. Ground truth annotations at 100 Hz provide both pupil center coordinates and blink/no-blink labels, enabling accurate performance assessment.

The neural network implementation uses PyTorch, with training and inference performed on an NVIDIA RTX A6000 GPU. We train for 200 epochs using single-sample batches to maintain temporal continuity. The loss function minimizes weighted RMSE between predicted and ground truth pupil positions. Training employs the Adam optimizer with an initial learning rate of 0.001. To prevent gradient instability in recurrent layers, we apply truncated backpropagation through time (TBPTT).

\subsubsection{Hardware Implementation}
The ASIC design targets GlobalFoundries 12LP-PLUS technology. We employ Cadence Genus for synthesis, Innovus for placement and routing, and Xcelium for verification. Power and performance metrics derive from post-layout simulations annotated with realistic switching activity patterns.

\subsection{Algorithm Performance}

\subsubsection{Architectural Optimization Analysis}
Table~\ref{table:optimization_ablation_study} quantifies the impact of each design decision in transforming ERVT into JaneEye-Net. Starting from the 150K-parameter baseline, our systematic optimizations achieve an 8.5$\times$ parameter reduction and 13.8$\times$ FLOPs reduction while improving accuracy.

The single-stage architecture unexpectedly improves accuracy from 2.53 to 2.37 pixels while reducing complexity by 38\%. This suggests that multi-scale processing, despite its success in general vision tasks, likely introduces unnecessary complexity for pupil tracking. Channel dimension adjustment yields the most dramatic efficiency gain, resulting in a 79.1\% reduction in FLOPs, although it temporarily degrades accuracy to 2.62 pixels. Subsequent optimizations recover this loss.

Our ConvJANET layer validates that simplified temporal modeling suffices for eye tracking. By eliminating input and output gates, we halve the computational cost of LSTM while maintaining comparable accuracy (2.41 vs. 2.53 pixels). The replacement of self-attention with convolution further reduces FLOPs by 58.5\% with minimal accuracy impact, confirming that local spatial relationships dominate in pupil detection.
\begin{table}[tbp]
\centering
\setlength{\tabcolsep}{2pt}
\renewcommand{\arraystretch}{1.3}
\begin{threeparttable}
\caption{Performance comparison between related works}
\label{table: software_performance_comparison}
\begin{tabular}{cccccc}
\toprule
\textbf{Model} & \textbf{\makecell{Event Frame\\Rate}} & \textbf{Parameters} & \textbf{FLOPs} & \textbf{Bit-width}\tnote{*} & \textbf{\makecell{Pixel\\Error}} \\
\midrule
Retina~\cite{bonazzi2024retina}  & $\leq$50kHz & 63K & 6.1M & W8A16 & 3.24\tnote{†} \\
MambaPupil~\cite{wang2024mambapupil}   & 20Hz & 8.59M & 2.61T & W32A32 & 2.03 \\
Go Sparse~\cite{zhang2024co}   & 20Hz & 178K & N/A & W8A8 & 3.71  \\
BigBrains~\cite{pei2024lightweight}  & 20Hz & 809K & 110.4M & W32A32 & 2.79  \\
ERVT~\cite{wang2024event}  & 20Hz & 150K & 74M & W32A32 & 2.48  \\
\textbf{Ours (Time)}  & 20Hz & \textbf{17.6K} & 10.7M & W8A16 & 2.45 \\
\textbf{Ours (Event count)} & 1.75--1250\,Hz & \textbf{17.6K} & 10.7M & W8A16 & 2.69 \\
\bottomrule
\end{tabular}
\begin{tablenotes}
\footnotesize
\item[*] W and A stand for weight and activation, respectively.
\item[†] All models use 3ET+ dataset except Retina, which uses Ini-30.
\end{tablenotes}
\end{threeparttable}
\end{table}
\subsubsection{Hardware-Software Co-Design Impact}
Table~\ref{table:software-hardware codesign} demonstrates the effectiveness of our progressive optimization strategy. While activation function replacements initially cause severe accuracy degradation (GELU$\rightarrow$ReLU: 6.28 pixels), targeted retraining recovers performance remarkably well. After all optimizations, including 8-bit weight and 16-bit activation quantization, total accuracy loss remains at just 1.6\% (from 2.41 to 2.45 pixels).
\begin{table*}[htbp]
\centering
\setlength{\tabcolsep}{0.6pt}
\renewcommand{\arraystretch}{1.4}
\begin{threeparttable}
\caption{Performance comparison of eye tracking systems}
\label{table: comparison}
\begin{tabular}{lcccccccccccc}
\toprule
\textbf{Work} &
\textbf{Dataset} &
\textbf{Prediction} & \textbf{Method} &
\textbf{Input Size} &
\textbf{Model Size} &
\textbf{Accuracy} & \textbf{\makecell{Frame Rate\\(FPS)}} &
\textbf{\makecell{Latency\\(ms)}} &
\textbf{\makecell{Hardware\\ Platform}} & \textbf{OP/s/W} & \textbf{\makecell{Energy/Frame\\($\mu$J)}} & \textbf{\makecell{EDP\\($\mu J\cdot\text{ms}$)}} \\
\midrule
{\cite{bonazzi2024retina}} & Ini-30 & Pupil & SNN & 64$\times$64$\times$2 & 63K & 3.24 px & $<$180 & 5.6 & Speck SoC & 158G\textsuperscript{†} & 16.1 & 89.7 \\
{\cite{zhang2024co}} & 3ET+ & Pupil & SCNN & 80$\times$60$\times$3 & 178K & 3.71 px & 1428 & 0.7 & FPGA SoC& N/A & 2,290 & 1603\\
{\cite{tan2025toward}} & EVBEYE & Gaze & DNN & 220$\times$120$\times$2 & 91K & 0.91$^\circ$(14.3 px) & 1252\textsuperscript{*} & $<$1\textsuperscript{*} & \makecell{ASIC \\@ 200 MHz} & 2560 & $<$12.7 & $<$12.7 \\
{\cite{zhao2022flatcam}} & OpenEDS & Gaze & MobileNetV2 & 96$\times$160$\times$1 & N/A & 3.16$^\circ$ & 253 & N/A & \makecell{ASIC \\@ 115 MHz} & N/A & 91.5 & N/A \\
{\cite{li2018etracker}} & Self-collected & Gaze & CNN & 224$\times$224$\times$3 & N/A & 0.54$^\circ$ & 60 & N/A & \makecell{GTX 745 \\@1020 MHz} & N/A & 9.2e5 & N/A \\
{\cite{feng2022real}} & OpenEDS & Gaze & DNN & N/A & 30K & 0.5$^\circ$ & 30 & N/A & \makecell{Jetson Xavier \\@1377 MHz} & N/A & 6.7e5 & N/A\\
{\cite{chen2022real}} & HE-Gaze & Gaze & GRU & 256$\times$192$\times$1 & N/A & 3.65$^\circ$ & 48 & 20.7 & \makecell{Snapdragon 845 \\@2.8 GHz} & N/A & 6.2e4 & 1.3e6\\
\textbf{Ours} & 3ET+ & Pupil & \textbf{JaneEye-Net} & 80$\times$60$\times$3 & \textbf{17.6K} & \textbf{2.45 px} & \textbf{2000} & \textbf{0.5} & \makecell{ASIC \\@ 400 MHz} & 567G & \textbf{18.9} & \textbf{9.5} \\
\bottomrule
\end{tabular}
\begin{tablenotes}
\footnotesize
\item[*]  1252 FPS is the frame rate derived from the sensor, and the hardware processing latency is reported to be $<$1ms.
\item[†] Calculated based on reported FLOPs, latency and power consumption.
\end{tablenotes}
\end{threeparttable}
\end{table*}
\subsubsection{Comparison with State-of-the-Art}
Table~\ref{table: software_performance_comparison} positions JaneEye-Net among existing event-based eye tracking methods. Our 17.6K-parameter model achieves several distinctions:

First, we demonstrate exceptional parameter efficiency evaluated on the same 3ET+ dataset, showcasing 8.5$\times$ fewer parameters than ERVT while achieving comparable accuracy (2.45 vs. 2.48 pixels). Second, our computational efficiency is 6.9$\times$ fewer \#FLOP than ERVT (10.7M vs. 74M) while maintaining competitive accuracy. This significant reduction is attributed to our architectural optimization, specifically the use of a parameter-efficient ConvJANET-based regression core and the removal of attention mechanisms.

Our event count-based variant adapts its frame rate from 1.75 Hz during fixation to 1250 Hz during saccades. This dynamic behavior naturally allocates computational resources based on the intensity of eye movements, improving both efficiency and tracking quality compared to fixed-rate approaches.

\subsection{Hardware Implementation Results}

\begin{figure}[t] 
\centering
\includegraphics[width=\linewidth]{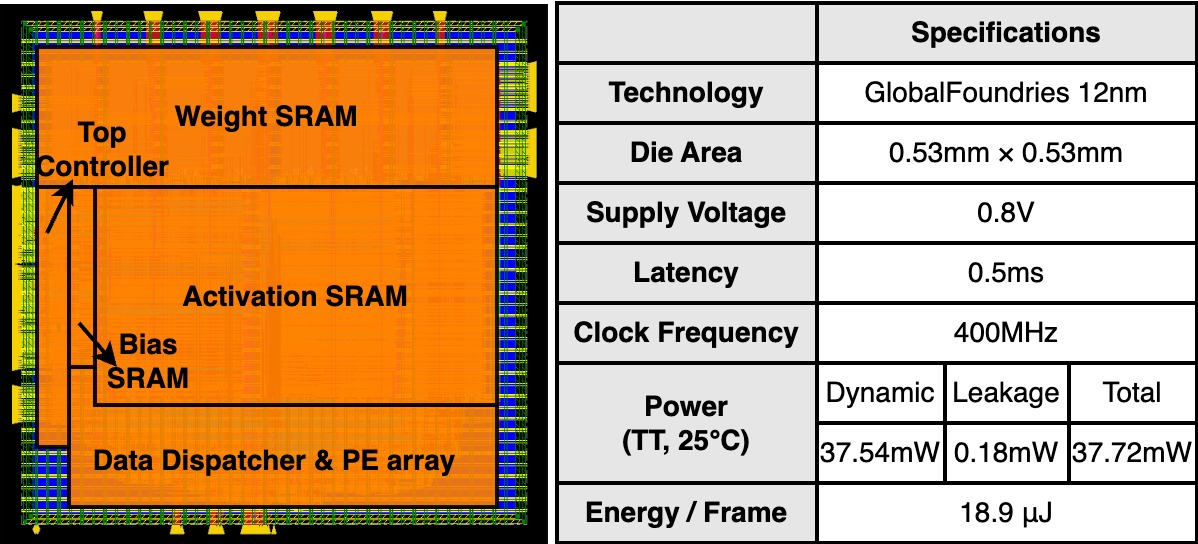}
\caption{Post-layout specification}
\label{Post-layout specification}
\end{figure}

Figure~\ref{Post-layout specification} summarizes the physical implementation characteristics. The design achieves 400 MHz operation at 0.8\,V in a compact 0.28\,mm$^2$ area (0.53$\times$0.53\,mm die). Power consumption totals 37.72\,mW, with dynamic power accounting for the majority (37.54\,mW) and minimal leakage (0.18\,mW), validating our low-voltage design strategy.

The accelerator can process each frame with a neural network inference latency of 0.5 ms @ 2000 FPS. Energy efficiency reaches 18.9 $\mu$J per frame, enabling extended operation on battery-powered devices. These metrics result from the co-design between our lightweight algorithm and optimized hardware architecture.

\subsubsection{System-Level Performance}
Table~\ref{table: comparison} provides a comprehensive comparison across eye tracking systems. JaneEye achieves the best Energy-Delay Product (EDP) of 9.5 $\mu$J$\cdot$ms among all systems, 26\% better than the previous ASIC implementation~\cite{tan2025toward}. This improvement combines three key factors: algorithmic efficiency (reduced required operations), architectural optimization (maximized data reuse), and technological advantages (12-nm vs. older process nodes).

Compared to GPU and mobile processor implementations, our ASIC demonstrates 35,450$\times$ better energy efficiency than Jetson Xavier and 3,280$\times$ better than Snapdragon 845, while maintaining competitive accuracy. In comparison to the only comparable ASIC~\cite{tan2025toward}, we achieve a 60\% higher frame rate with only 48\% more energy consumption, resulting in superior overall efficiency.

The 567 GOP/s/W computational efficiency enables real-time processing within tight power budgets, making JaneEye suitable for integration into next-generation XR wearables where battery life remains a critical constraint.

Table~\ref{table: comparison} presents a comparative analysis of various eye tracking systems. The proposed eye tracking system in this work achieves the highest accuracy in the task of pupil detection, and it outperforms all referenced methods in terms of speed at 2000 FPS and energy–delay product (EDP) 9.5 $\mu$J$\cdot$ms, achieving the best overall performance–efficiency trade-off. Our design also achieves relatively low energy consumption with only 18.9 $\mu$J/Frame. Compared to the state-of-the-art eye-tracking ASIC presented in~\cite{tan2025toward}, which targets gaze detection, our system achieves a 60\% higher frame rate while incurring only a 48\% increase in energy consumption, resulting in a 26\% EDP reduction. This highlights the superior performance-efficiency trade-off of our work.

\section{Conclusion}
This paper presents JaneEye, a breakthrough event-based eye tracking system addressing critical XR requirements. We developed an ultra-lightweight neural network that achieves 2.45-pixel accuracy with only 17.6K parameters, utilizing our novel ConvJANET architecture. Our co-designed 12-nm ASIC accelerator delivers exceptional performance, achieving 2000 FPS with 0.5 ms latency and an energy consumption of 18.9 $\mu$J/frame. Given efficiency demands for wearable integration, our implementation achieves superior EDP while maintaining competitive accuracy. This work establishes new benchmarks for neuromorphic computing in resource-constrained environments, enabling practical deployment of high-performance eye tracking in next-generation immersive technologies.

\section*{Acknowledgement}
We thank GlobalFoundries for providing us the 12LP-PLUS PDK through the GF12+ University Partnership Programme. This work was partially supported by the Dutch Research Council (NWO) under the Talent Programme Veni 2023 scheme in Applied and Engineering Sciences (AES), Grant No. 21132 (Energy-Efficient Real-Time Edge Intelligence for Wearable Healthcare Devices).

\bibliographystyle{IEEEtran}
\bibliography{IEEEabrv,ref} 

\end{document}